\documentclass[prd,twocolumn,a4paper,superscriptaddress]{revtex4}
\usepackage{graphicx}
\usepackage{amsmath,amsfonts,amstext,amssymb,amsbsy,amsopn,amsthm,eucal}
\usepackage{dsfont}

\begin{document}

\newcommand{\be}{\begin{equation}}
\newcommand{\ee}{\end{equation}}
\newcommand{\bq}{\begin{eqnarray}}
\newcommand{\eq}{\end{eqnarray}}
\newcommand{\bsq}{\begin{subequations}}
\newcommand{\esq}{\end{subequations}}
\newcommand{\bc}{\begin{center}}
\newcommand{\ec}{\end{center}}
\newcommand\lsim{\mathrel{\rlap{\lower4pt\hbox{\hskip1pt$\sim$}}
    \raise1pt\hbox{$<$}}}
\newcommand\gsim{\mathrel{\rlap{\lower4pt\hbox{\hskip1pt$\sim$}}
    \raise1pt\hbox{$>$}}}
\newcommand{\nn}{\nonumber}
\newcommand{\bo}{\raise-1mm\hbox{\Large$\Box$}} 
\newcommand{\simpropto}{
\begin{array}{c}
\propto \\[-1.6ex] \sim
\end{array}}

\title{Mass inflation in a $D$ dimensional Reissner-Nordstr\"om black hole: \\  a hierarchy of particle accelerators ?}
\author{P.P. Avelino}
\email[Electronic address: ]{ppavelin@fc.up.pt}
\affiliation{Centro de Astrof\'\i sica da Universidade do Porto, Rua das Estrelas, 4150-762 Porto, Portugal,}
\affiliation{Departamento de F\'{\i}sica da Faculdade de Ci\^encias
da Universidade do Porto and Centro de F\'{\i}sica do Porto, \\  
Rua do Campo Alegre 687, 4169-007 Porto, Portugal}
\author{A.J.S. Hamilton}
\email[Electronic address: ]{Andrew.Hamilton@colorado.edu}
\affiliation{JILA and Dept. Astrophysical $\&$ Planetary Sciences, \\
Box 440, U. Colorado, Boulder, CO 80309, USA}
\author{C.A.R. Herdeiro}
\email[Electronic address: ]{herdeiro@ua.pt}
\affiliation{Departamento de F\'{\i}sica da Universidade de Aveiro \& I3N, \\ 
Campus de Santiago, 3810-193 Aveiro, Portugal}
\author{M. Zilh\~ao$^2$}
\email[Electronic address: ]{mzilhao@fc.up.pt}
\date{May 2011}
\begin{abstract}
 We study the geometry inside the event horizon of perturbed $D$ dimensional Reissner-Nordstr\"om-(A)dS type black holes showing that,  similarly to the four dimensional case, mass inflation also occurs for $D>4$.  First, using the homogeneous approximation, we show that an increase of the number of spatial dimensions contributes to a steeper variation of the metric coefficients with the areal radius and that the phenomenon is insensitive to the cosmological constant in leading order. Then, using the code reported in \cite{Avelino:2009vv} adapted to $D$ dimensions, we perform fully non-linear numerical simulations. We perturb the black hole with a compact pulse adapting the pulse amplitude such that the relative variation of the black hole mass is the same in all dimensions, and determine how the black hole interior evolves under the perturbation. We qualitatively confirm that the phenomenon is similar to four dimensions as well as the behaviour observed in the homogeneous approximation. We speculate about the formation of black holes inside black holes triggered by mass inflation, and about possible consequences of this scenario.
\end{abstract}

\maketitle

\section{\label{intr}Introduction}
It is well established that the inner horizon of the Reissner-Nordstr\"om  (RN) black hole is unstable and evolves into a curvature singularity when perturbed \cite{Penrose:1968ar,Chandra,Poisson:1989zz,Poisson:1990eh,Burko:1997zy,Burko:1997tb,Burko:1998jz,Burko:2002qr,Burko:2002fv,Oren:2003gp,Hansen:2005am,Dafermos:2003wr,Hamilton:2008zz,Avelino:2009vv}. The formation of this singularity is a consequence of a large relativistic counter-streaming produced by the perturbation in the vicinity of the inner horizon. Thus the black hole interior, when perturbed,  effectively becomes a particle accelerator, with centre of mass energy swiftly reaching Planckian regimes. One immediate consequence is  that Planckian and trans-Planckian physics becomes relevant inside a black hole before the curvature scale of the unperturbed black hole becomes Planckian (in the latter case this should only occur close to the central curvature singularity of the classical solution). Trans-Planckian scattering is dominated by graviton exchange \cite{'tHooft:1987rb} and, for sufficiently small impact parameter, black holes should form \cite{Choptuik:2009ww}. Thus, the above picture of mass inflation leads us to the fascinating scenario where \textit{black holes could form inside black holes.}

 In recent years, the physics of trans-Planckian scattering has been intensively explored motivated by models of large \cite{ArkaniHamed:1998rs,Antoniadis:1998ig,ArkaniHamed:1998nn} or infinite \cite{Randall:1999ee,Randall:1999vf} extra dimensions. In these scenarios the fundamental Planck scale could be considerably smaller than the apparent four dimensional Planck scale ($10^{19}$ GeV) and black holes could even potentially be formed in ongoing particle accelerator runs \cite{Dimopoulos:2001hw,Khachatryan:2010wx}. These black holes would be higher dimensional charged black holes, by virtue of the charge of the colliding partons. Then, if these higher dimensional black holes also exhibit the same type of mass inflation instability as their four dimensional counterparts, we could have the amusing scenario of a (black hole) \textit{particle accelerator inside a} (terrestrial) \textit{particle accelerator}.

In this paper we show that higher dimensional charged black holes exhibit mass inflation, quite analogous to their four dimensional counterparts (see \cite{Hwang:2011mn} for a recent study of lower dimensional black holes). More concretely we consider dynamical charged black hole solutions which generalise the standard Reissner-Nordstr\"om solution in $D$ dimensional Einstein-Maxwell theory coupled to a cosmological constant (Sections \ref{fequations} -- \ref{anstaznc}). The role of the latter is shown to be quite irrelevant by considering the homogeneous approximation (Section \ref{homoap}), which also allows the observation that the phenomenon is faster (in terms of the areal radius) as one increases the dimension, for the same (perturbation) energy density placed at the same areal radius. We then perform full numerical simulations of a compact wave packet falling into the black hole (Section \ref{nl}), using the code described in \cite{Avelino:2009vv}. To make a comparison between the different dimensions, we perform runs in $D=4,5,6,7$ varying the amplitude of the pulse in such a way that the relative black hole mass variation is constant, approximately $25\%$. We exhibit the dynamics of the inner (and also of the outer) horizon, which is qualitatively quite similar in all dimensions. We also display the behaviour of the mass function and areal radius inside the horizon, from which a qualitative agreement with the observations made for the homogeneous approximation may be inferred.

\section{\label{fequations} Action and Field equations}

Consider the Einstein-Maxwell action with a cosmological constant and minimally coupled to a scalar field in a $D$ dimensional spacetime:
\be
S=\int d^D x {\sqrt {-g}} \left(R - \frac{(D-1)(D-2)}{3}\Lambda  + 16 \pi {\cal L}_M \right) \ , 
\label{action}
\ee
where $R$ is the Ricci scalar and $\Lambda$ is the cosmological constant. ${\cal L}_M$ is the matter Lagrangian, which we will assume 
to be the sum of the usual Maxwell contribution, ${\cal L}_F$, and the 
contribution from a self-gravitating real massless scalar field $\varphi$: 
\be
{\cal L}_M = {\cal L}_F + {\cal L}_\varphi = - \frac{F^2}{16\pi} - \frac{1}{8 \pi} 
\varphi_{,\alpha} \varphi^{,\alpha}\, \ , \label{matterlag}
\ee
where $F^2=F_{\alpha \beta} F^{\alpha \beta}$ and $F_{\alpha \beta}$ is the Maxwell tensor. Our signature choice 
is $-+++\dots$. 
The energy-momentum tensor of the matter fields is given by
 $ {}^F T_{\mu \nu} + {}^\varphi T_{\mu \nu}$, where
\bq
{}^F T_{\mu \nu} &=& \frac{1}{4\pi}\left(F_{\mu \alpha} {F_\nu}^\alpha - \frac{1}{4} g_{\mu \nu} F^2\right)\ , \\
{}^\varphi T_{\mu \nu} &=& \frac{1}{4 \pi} \left(\varphi_{,\mu} \varphi_{,\nu} - \frac{1}{2} g_{\mu \nu} \varphi_{,\alpha} \varphi^{,\alpha}\right)\, \ .
\eq
The equations of motion derived from (\ref{action}) with (\ref{matterlag}) are the gravitational equations
\bq
G_{\mu \nu} +\frac{(D-1)(D-2)}{6}\Lambda g_{\mu\nu}&=& 8 \pi ( {}^F T_{\mu \nu} + {}^\varphi T_{\mu \nu})  \, \ , \label{einstein}
\eq
the matter equations
\be
\label{eqphif}
\Box \varphi=0\, \ , \qquad d\star F=0 \ , 
\ee
where $\star$ denotes Hodge dual.

\section{$D$ dimensional Reissner-Nordstr\"om-(A)dS}
\label{rnsolution}
A well known solution to this system is that of a $D$ dimensional spherically symmetric (i.e. with spatial isometry $SO(D-1)$) charged black hole in asymptotically (Anti)-de Sitter spacetime, first discussed by Tangherlini \cite{Tangherlini:1963bw} (see also \cite{Cardoso:2004uz}). The fields read
\be
ds^2=-f(r)dt^2+\frac{dr^2}{f(r)}+r^2d\Omega_{D-2}  \label{rnmetric1}\ , \ee
\be
F=-\frac{Q}{r^{D-2}}dt\wedge dr \ , \qquad \varphi=0 \ , \ee
where $d\Omega_{D-2}$ is the line element on the $(D-2)$-sphere and 
\be
f(r)=1-\frac{2M}{r^{D-3}}+\frac{q^2}{r^{2(D-3)}}-\frac{\Lambda}{3}r^2 \label{rnmetric2}\ . \ee
We have chosen the coefficient of the cosmological constant term in the gravitational action such that the $\Lambda$ term in the last equation is dimension independent. The two metric parameters $M$ and $q$ are related to the ADM mass and charge by
\be
M_{ADM}=\frac{(D-2)\mathcal{A}_{D-2}}{8\pi}M \ , \ee
\be
Q=\sqrt{\frac{(D-3)(D-2)}{2}}q \ , \label{Q}\ee
where $\mathcal{A}_{D-2}$ is the area of a $(D-2)$-sphere of unit radius.

The solution has Killing horizons of the $\partial/\partial t$ Killing vector field at radial distances $r$ obeying
\be
r^{2(D-3)}-2Mr^{D-3}+q^2-\frac{\Lambda}{3} r^{2(D-2)}=0 \ . \label{horizons}\ee
If $\Lambda=0$, the solutions are a simple generalisation of the four dimensional case; the horizons are located at
\be
r_{\pm}=\left[M\pm \sqrt{M^2-q^2}\right]^{\frac{1}{D-3}} \ . \label{rpm}\ee
Thus, the higher dimensional Reissner-Nordstr\"om-Tangherlini black hole still has an inner, at $r=r_-$, and an outer horizon, at $r=r_+$, as long as $M\ge |q|$. In this paper we shall always take $M=1$ and $q=0.95$ for the unperturbed black hole.

If $\Lambda\neq 0$ one cannot, in general, provide an analytical solution for (\ref{horizons}). Qualitatively, if $\Lambda<0$, the spacetime is asymptotically Anti-de-Sitter and there are no further Killing horizons; if $\Lambda>0$, the spacetime is asymptotically de-Sitter and there may be further Killing horizons of cosmological nature.

\section{Ansatz with spherical symmetry in double null coordinates}
\label{anstaznc}
To solve numerically the coupled system  (\ref{einstein}) and (\ref{eqphif}) is, in general, quite a difficult task. In fact, numerical relativity in higher dimensions is a very recent field (see eg. \cite{Yoshino:2009xp,Shibata:2009ad,Zilhao:2010sr,Shibata:2010wz,Witek:2010xi,Witek:2010az,Okawa:2011fv} for general formulations with applications). Here we are interested in spherically symmetric black holes, perturbed in such a way that this symmetry is preserved. This makes the system tractable. Thus, we take a spherically symmetric ansatz written in  double-null coordinates:
\be
ds^2=-2e^{2\sigma(u,v)}du dv + r^2(u,v) d\Omega_{D-2}\, \ , \label{metric}
\ee
\be
F=F_{uv}(u,v)du\wedge dv \ , \qquad  \varphi=\varphi(u,v) \  , \ee
where $u$ and $v$ are taken to be ingoing and outgoing respectively. 
For the unperturbed solution of the previous section, $\varphi=0$, $e^{2\sigma(u,v)}=f(r)$ with
\be
u=\frac{1}{\sqrt{2}}(t-r^*) \ , \qquad v=\frac{1}{\sqrt{2}}(t+r^*) \ , \ee
and $dr/dr^*=f(r)$.

The Maxwell equations~(\ref{eqphif}) are simply solved to yield
\be
\label{maxwell1}
F_{u v}r^{D-2}e^{-2\sigma}= {\rm constant}= Q \, \ ,
\ee
where
\be Q=\frac{1}{2\mathcal{A}_{D-2}}\oint F \ , \ee
is the electric charge. The electric field is therefore purely radial, as expected from spherical symmetry. The scalar field equation~(\ref{eqphif}) gives
\be
\varphi_{,uv} = -\frac{D-2}{2r}\left(r_{,v} \varphi_{,u} + 
r_{,u} \varphi_{,v}\right) \ .
\label{scalar}
\ee

The $uu$, $vv$, $uv$ and transverse components of the Einstein equations~(\ref{einstein}), which are the only distinct non-trivial ones, give rise to the following equations
(the left hand sides of these equations are components of the Einstein tensor
$G_{\mu\nu}$):
\be  \frac{2 r_{,u} \sigma_{,u} -  r_{,uu}}{r}=\frac{2}{D-2}(\varphi_{,u})^2  \ , \label{einuu}\ee
\be  \frac{2 r_{,v} \sigma_{,v} -  r_{,vv}}{r}=\frac{2}{D-2}(\varphi_{,v})^2  \ , \label{einvv}\ee
\begin{eqnarray} 
\frac{(D-3)(e^{2 \sigma} + 2 r_{,v} r_{,u}) +2 r r_{,uv}}{2r^2e^{2 \sigma} }= \nonumber \\
= \frac{Q^2}{(D-2)r^{2(D-2)}}+\frac{D-1}{6}\Lambda  \ , \ \ \  \ \ \label{einuv} \end{eqnarray}
\begin{eqnarray}
-(D-3)\left\{(D-4)\left[\frac{e^{2\sigma}}{2}+r_{,u}r_{,v}\right]+2  r r_{,uv}\right\}-2r^2\sigma_{,uv} = \nn \\ =e^{2 \sigma}\left( \frac{Q^2}{ r^{2(D-3)}}-\frac{(D-1)(D-2)}{6}\Lambda r^2\right)+2r^2\varphi_{,u}\varphi_{,v} \ . \ \ \ \ \label{eintt}  \end{eqnarray}
For $D=4$, $\Lambda=0$, these equations reduce to the ones in \cite{Avelino:2009vv} with constant Brans-Dicke scalar.

Equations~\eqref{einuv}-\eqref{eintt} are the evolution equations, that determine the dynamical variables  in the future of the hypersurfaces where initial conditions are set (see Fig. 1 in  \cite{Avelino:2009vv}). 
In equations~\eqref{einuu}-\eqref{einvv}, the contribution from the energy-momentum of the scalar field is
\be T^{\varphi}_{uu}=\frac{(\varphi_{,u})^2}{4\pi} \ , \qquad T^{\varphi}_{vv}=\frac{(\varphi_{,v})^2}{4\pi} \ , \ee
respectively, which represent, physically, the flux of the scalar field through surfaces of constant $v$ and $u$, i.e.\ outflux and influx. We shall impose that, initially, only influx exists. However, outflux is inevitably produced by scattering off the spacetime geometry.

\section{Homogeneous approximation}
\label{homoap}
In the homogeneous approximation one considers only $r$ dependence in the line element (see \cite{Avelino:2009vv} for a discussion of this approximation and references).  That is we keep the symmetry with respect to the $\partial / \partial t$ Killing vector, which is spacelike in between the inner and outer horizons of the unperturbed solution. The spherically symmetric homogeneous line element is given by
\be
ds^2=g_{tt}(r)dt^2+g_{rr}(r)dr^2+r^2d\Omega_{D-2} \ . \ee
The black hole charge will produce a purely radial electrostatic field. The Maxwell equations then yield
\be
F_{tr}=-\frac{Q}{r^{D-2}}\sqrt{-g_{tt}g_{rr}} \ . \ee
The scalar field admits solely a radial dependence, and the Klein-Gordon equation may be written as a first order equation
\be
\varphi' \propto \sqrt{\left|\frac{g_{rr}}{g_{tt}}\right|}\frac{1}{r^{D-2}} \ , \ee
where the prime denotes radial derivative. The $tt$ and $rr$ components of the Einstein equations read, respectively
\begin{eqnarray} 
\frac{D-2}{2r^2g_{rr}}\left(rg_{rr}'+(D-3)g_{rr}(g_{rr}-1)-\frac{\Lambda }{3}(D-1)r^2g_{rr}^2\right) \nn \\ =\frac{Q^2g_{rr}}{r^{2(D-2)}}+\varphi'^2  \ ,\ \ \ \ \ \label{eqhomo1}\end{eqnarray}
\begin{eqnarray} 
\frac{D-2}{2r^2g_{tt}}\left(rg_{tt}'-(D-3)g_{tt}(g_{rr}-1)+\frac{\Lambda }{3}(D-1)r^2g_{rr}g_{tt}\right) \nn \\=-\frac{Q^2g_{rr}}{r^{2(D-2)}}+\varphi'^2  \ . \ \ \ \ \ \label{eqhomo2}\end{eqnarray}

\subsection{Analytic analysis for the homogeneous approximation}
\begin{figure}[h!tb]
  \centering
    \includegraphics[clip=true,width=0.46\textwidth]{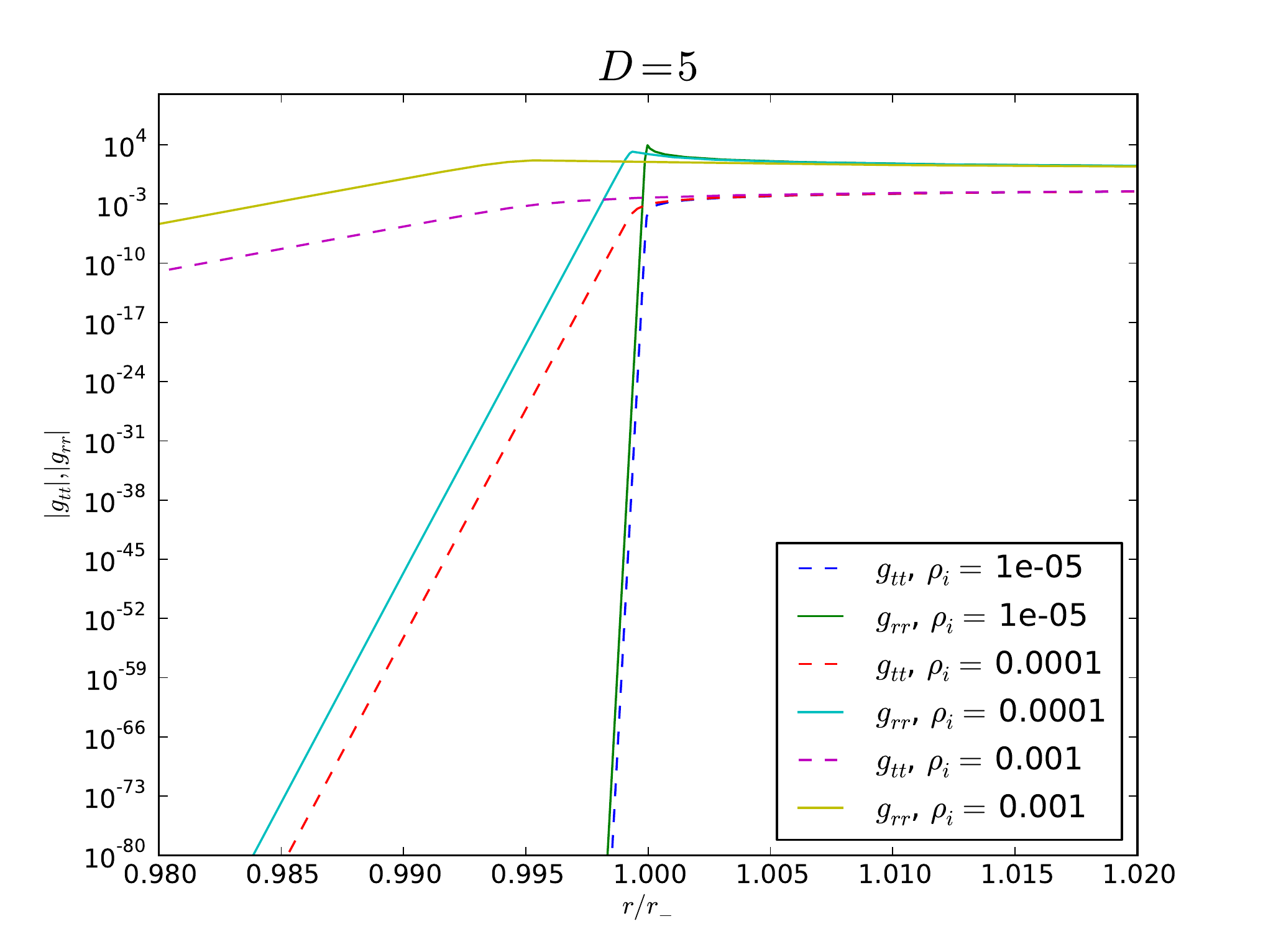}
  \caption{Variation of the radial and temporal metric components with the areal radius normalised by the value at the inner horizon, for a $D=5$ charged black hole.}
  \label{d5}
\end{figure}

\begin{figure}[h!tb]
  \centering
    \includegraphics[clip=true,width=0.46\textwidth]{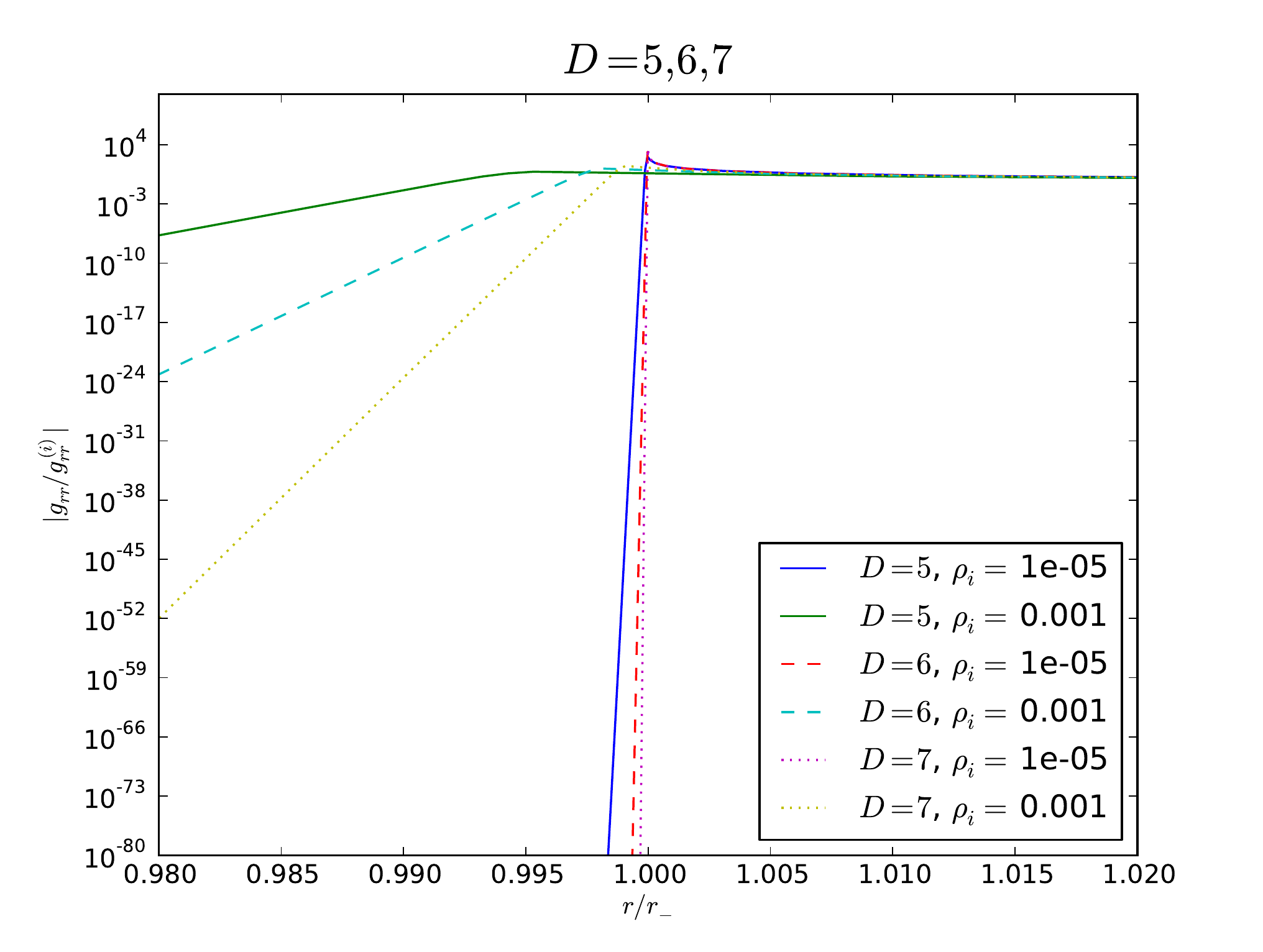}
  \caption{Comparison of the radial metric component behaviour for $D=5,6,7$ and for two different values of $\rho_i$, in terms of the areal radius normalised by the value at the inner horizon.}
  \label{d678}
\end{figure}

The scalar field we have been considering may be regarded as a perfect fluid (see eg. \cite{Babichev:2008dy}) with
\be
p=\rho=4\pi \mathcal{L}_\varphi \ . \ee
In the homogeneous approximation $\varphi=\varphi(r)$. Then, energy momentum conservation for the scalar field yields
\be
\rho=\rho_i\frac{g_{tti}}{g_{tt}}\left(\frac{r_i}{r}\right)^{2(D-2)} \ , \label{rho} \ee
where $\rho_i$, $g_{tti}$ and $r_i$ are all integration constants; $\rho_i$ is the density at the surface $r=r_i$, where $g_{tt}$ is $g_{tti}$.
In the mass inflation region, we expect the variations of $g_{tt}$ and $g_{rr}$ to be fast. Thus, we approximate (\ref{eqhomo1}) and (\ref{eqhomo2}) by, respectively
\be
\frac{D-2}{2r}\frac{g_{rr}'}{g_{rr}}\simeq \varphi'^2 \ , \ee
\be
\frac{D-2}{2r}\frac{g_{tt}'}{g_{tt}}\simeq \varphi'^2 \ .  \label{ap1}\ee
Observe that the cosmological constant becomes irrelevant, in leading order. Moreover, we conclude that, as in four dimensions, the radial and temporal components of the metric should be approximately proportional
\be 
-g_{rr} \propto g_{tt} \ .\label{grrproptogtt}\ee
It follows from this result, together with (\ref{rho}) and (\ref{ap1}) that 
\be
\frac{g_{rr}'}{g_{rr}}\propto \frac{1}{r^{2D-5}} \ \Rightarrow \ \ g_{rr}\propto e^{-\left(\frac{R}{r}\right)^{2(D-3)}} \ ,   \ee
where
\be
R \simeq r_i \left(\frac{-2 \rho_i r_i^2 g_{tti}}{(D-2)(D-3)} \frac{g_{rr}}{g_{tt}}\right)^{1/(2(D-3))}  \label{R}\ ,   \ee
is roughly constant in the mass inflation region. The transition from the RN phase (where $g_{rr}$ and $g_{tt}$ given approximately by Eqs. (\ref{rnmetric1}) and (\ref{rnmetric2}) with $-g_{rr} g_{tt}=1$) to the mass inflation phase (where Eq. (\ref{grrproptogtt}) is approximately valid) is rather sharp and can be defined by the equality of the two terms on the r.h.s. of Eqs.  (\ref{eqhomo1}) and (\ref{eqhomo2}). Taking into account that at the transition $(-g_{rr}/g_{tt})^{1/2} \simeq -g_{rr} \simeq 1/g_{tt}$, one finds that, in the mass inflation region,
\be
-\frac{g_{rr}}{g_{tt}}= \frac{Q^4}{4 \rho_i^2g_{tti}^2 r_i^{4(D-2)}} \label{grrogtt} \ .   \ee
Taking the logarithm of $g_{rr}$ and expanding it up to linear order in $r-r_-$ in the mass inflation region ($r \lsim r_-$) one gets
\be 
\ln g_{rr} \simeq 2(D-3)\left(\frac{R}{r_-}\right)^{2(D-3)}\frac{r}{r_-}+{\rm const}\,. \label{lngrr} \ee
Using Eqs. (\ref{Q}), (\ref{rpm}), (\ref{R}), (\ref{grrogtt}) and (\ref{lngrr}) one finally obtains 
\be 
\ln g_{rr} \simeq f(M,q) \frac{(D-3)^2(D-2)}{\rho_i r_i^{2(D-2)} g_{tti}}\frac{r}{r_-}+{\rm const}\,, \label{lngrr1} \ee
with
\be 
f(M,q)=\frac{q^4}{4(M-{\sqrt{M^2-q^2}})^2}\,. \ee
This leads to the near straight lines in the mass inflation region in Fig. \ref{d5}, where we plot the variation of the radial and temporal metric components for $D=5$. As in the four dimensional case, Fig. \ref{d5} shows that mass inflation is more abrupt for smaller values of $\rho_i$. This happens because $d\ln g_{rr}/dr \propto 1/\rho_i$ in the mass inflation region which results in a larger slope of $\ln g_{rr}(r)$ for smaller values of $\rho_i$. In higher dimensions the behaviour is similar but the relative variation of the radial metric component is more abrupt as the spacetime dimension increases. Fixing $r_i$ and $\rho_i$, $d\ln g_{rr}/dr \propto (D-3)^2(D-2)$ if $r_i=1$. This can be confirmed in Fig. \ref{d678}, where we plot the radial metric component for two different values of $\rho_i$ and for $D=5,6,7$. Indeed Fig. \ref{d678}  shows that the slopes increase with $D$, for a fixed $\rho_i$. Note that the radial coordinate is the areal radius, therefore having an invariant geometric meaning. Thus it is the appropriate variable to compare the mass inflation phenomenon in different dimensions. In order to make this comparison, however, we fixed $r_i=1$ in all dimensions. In the next section, we shall fix the same physical observable, namely the same relative mass variation for all dimensions.
 
\section{Full numerical simulations}
\label{nl}

\begin{figure}[h!tb]
  \centering
    \includegraphics[clip=true,width=0.53\textwidth]{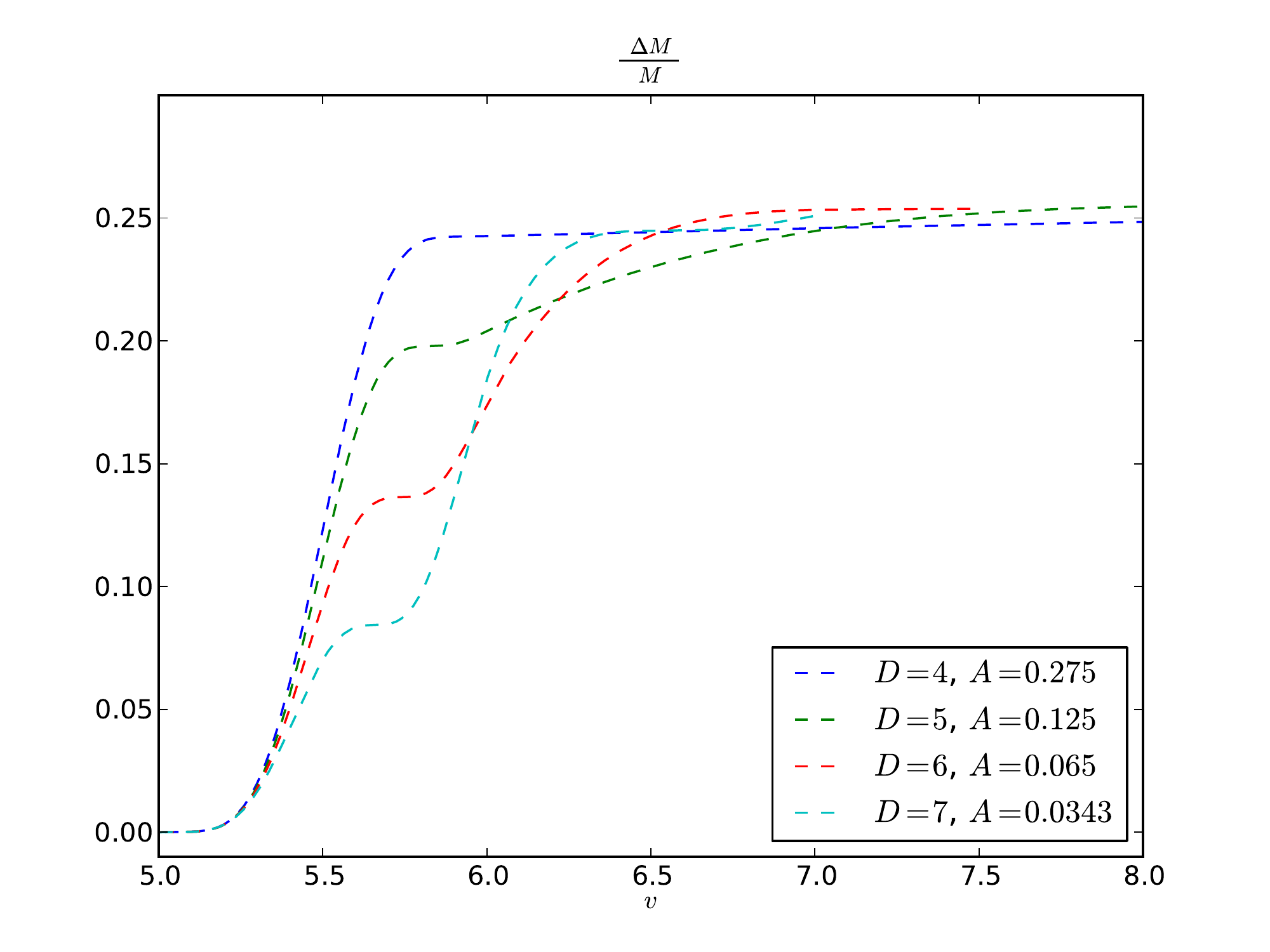}
  \caption{Evolution of the black hole mass as it interacts with the pulse. The amplitude was chosen, in each dimension, so that the final black hole mass variation is approximately $25\%$ for all cases.}
  \label{mass_global}
\end{figure}

We shall now present the results we have obtained using the numerical code reported in \cite{Avelino:2009vv} generalised to $D$ dimensions and the compact pulse used therein, which is,  along the initial null segment $u = u_0$,
\begin{equation}
\varphi_{,v}(u_0,v)=A \sin^2\left(\pi \frac{v-v_0}{\Delta v}\right)
\label{pulse}\ , \ \ v_0\le v \le v_0+\Delta v \ ,
\end{equation} 
and vanishing outside the interval $v_0$ to $v_0 + \Delta v$, with an interval $  \Delta v = 1$. With this code we have performed fully non-linear numerical simulations in $D=4,5,6,7$. In order to make a meaningful comparison between the different dimensions, our strategy was to fix the same relative mass variation, $\Delta M/M$,  after the interaction with the pulse, for the different dimensions. In order to achieve this we had to vary the amplitude of the pulse $A$ by an iterative procedure. In Fig. \ref{mass_global} we show the evolution of the mass of the black hole for the values of the amplitude that were found to give, for each dimension, a relative mass variation of about $25\%$.

\begin{figure}[h!tb]
  \centering
    \includegraphics[clip=true,width=0.53\textwidth]{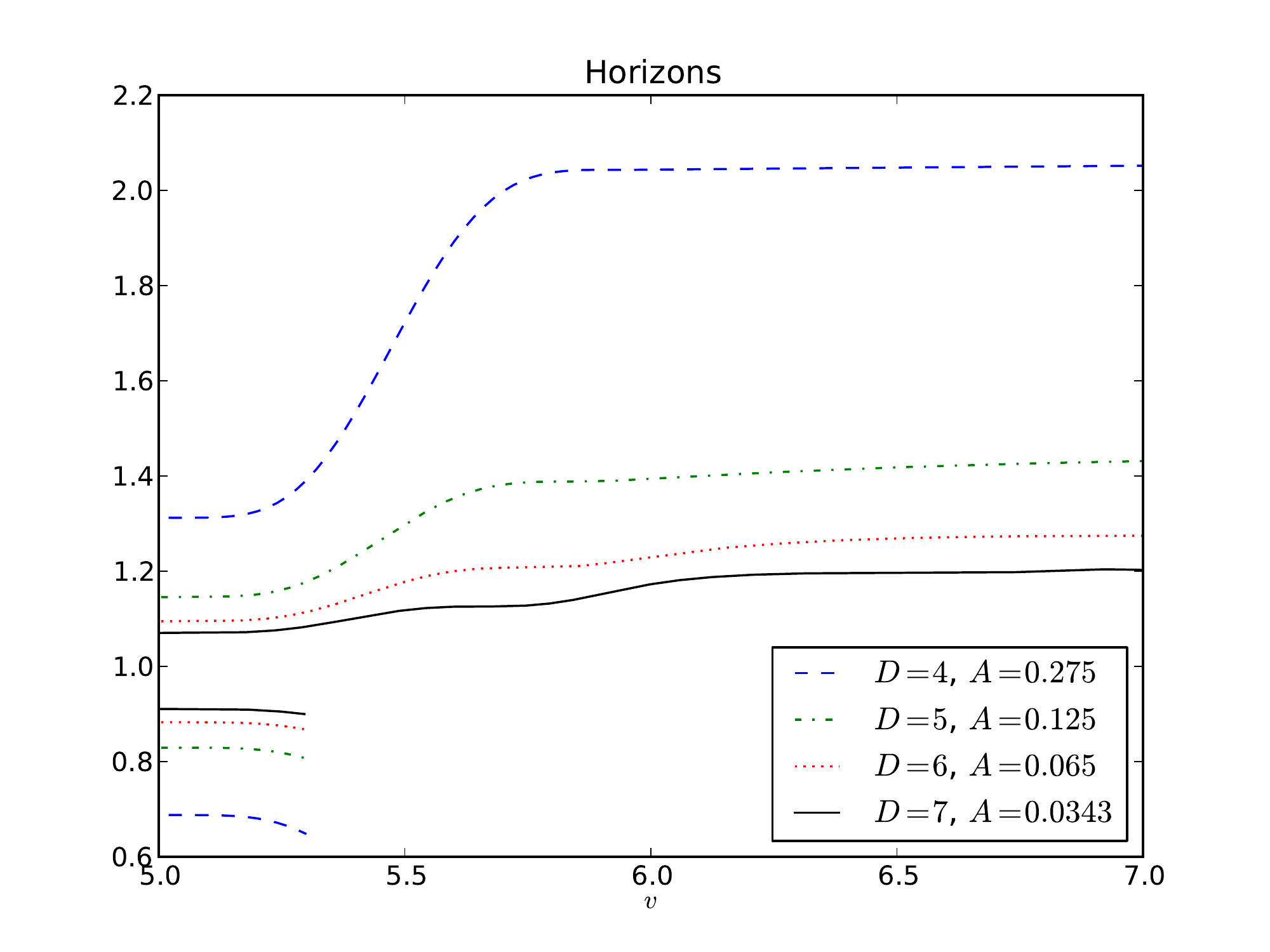}
  \caption{Evolution of the black hole apparent horizons. For each dimension the outer (top curve) and inner (bottom curve) are plotted.}
  \label{horizon5}
\end{figure}

In order to follow the black hole mass, we computed the (outermost) apparent horizon at each null slice, i.e. the set of points $(u_a,v_a)$ such that
\begin{equation} \frac{\partial r}{\partial v}(u_a,v_a)=0  \ . \label{ahorizon}
\end{equation}
In Fig. \ref{horizon5} we exhibit the apparent horizon evolution (both inner and outer). Then, the black hole mass was given by the Misner-Sharpe \cite{Misner} mass function evaluated at this apparent horizon. The Misner-Sharpe mass is the total effective mass inside a sphere of radius $r(u,v)$, and is equal to
\bq
M(u,v)&=&\frac{r^{D-3}}{2}\left(1+\frac{q^2}{r^{2(D-3)}}-g_{rr}^{-1}\right)  \nn \\ 
&=&\frac{r^{D-3}}{2}\left(1+\frac{q^2}{r^{2(D-3)}}+4\frac{r_{,u} r_{,v}}{2 e^{2\sigma}}\right)\label{massinf}\,.
\eq

\begin{figure*}[h!tb]
  \centering
    \includegraphics[clip=true,width=0.49\textwidth]{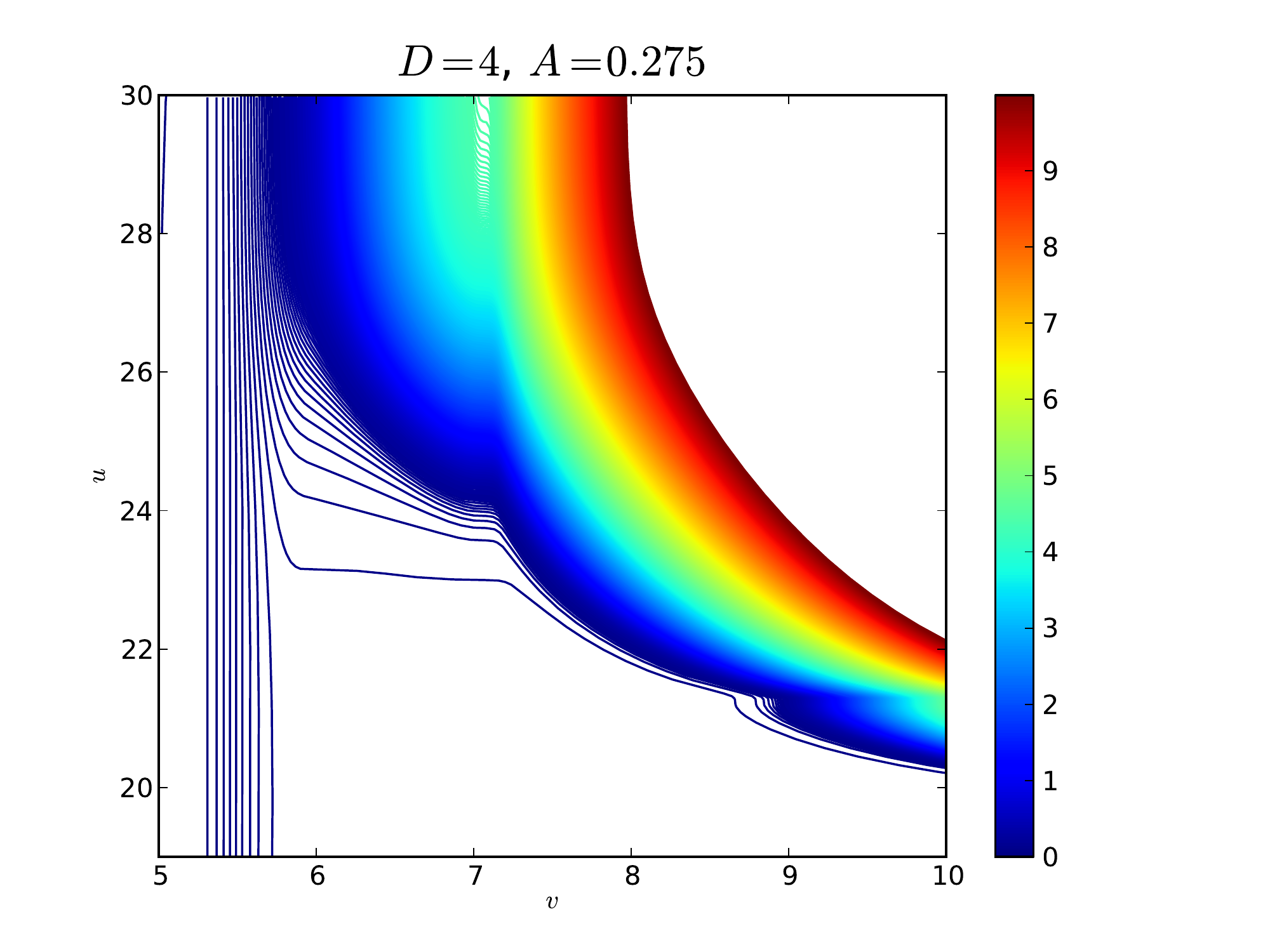}
      \includegraphics[clip=true,width=0.49\textwidth]{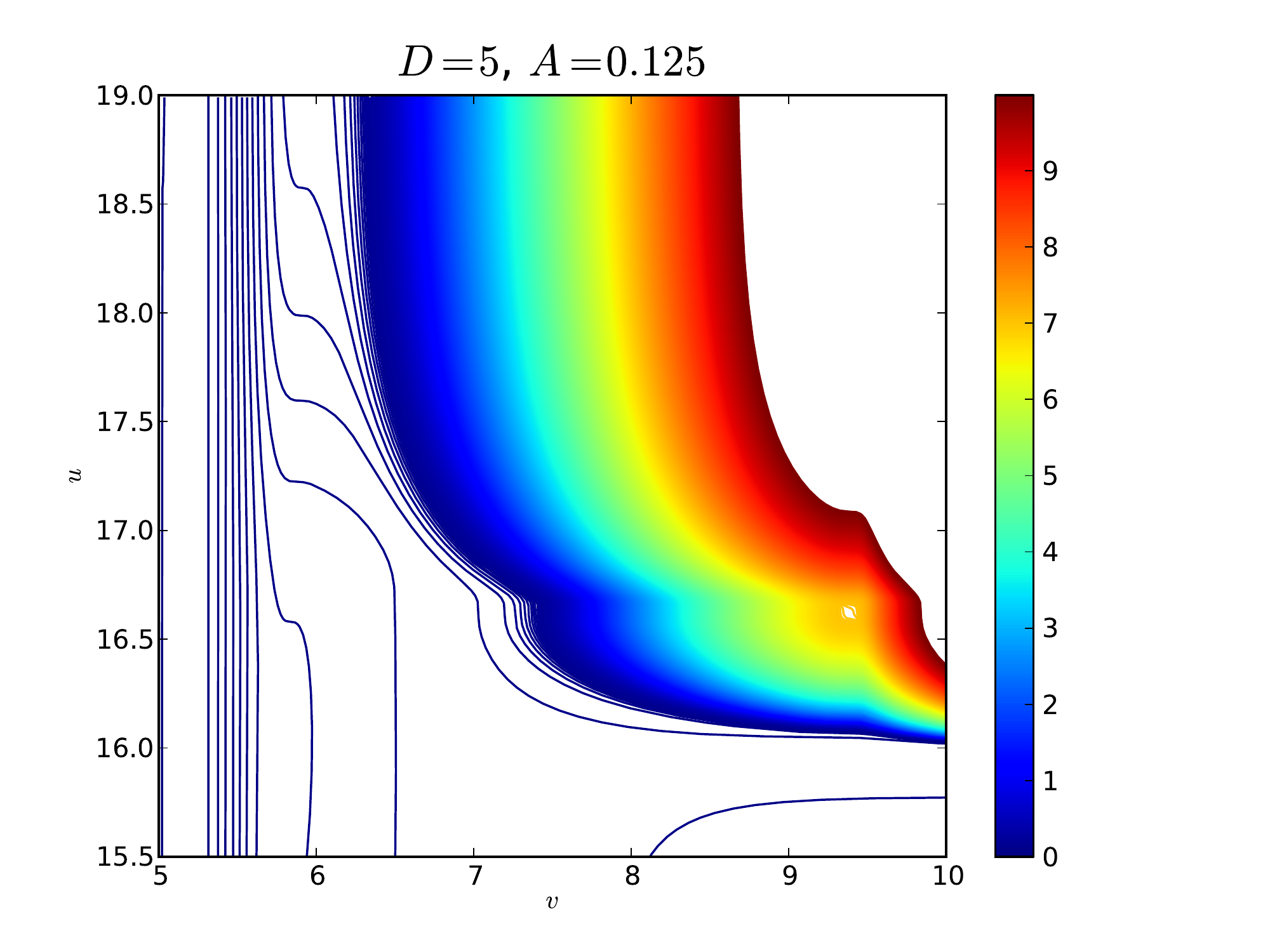}
       \includegraphics[clip=true,width=0.49\textwidth]{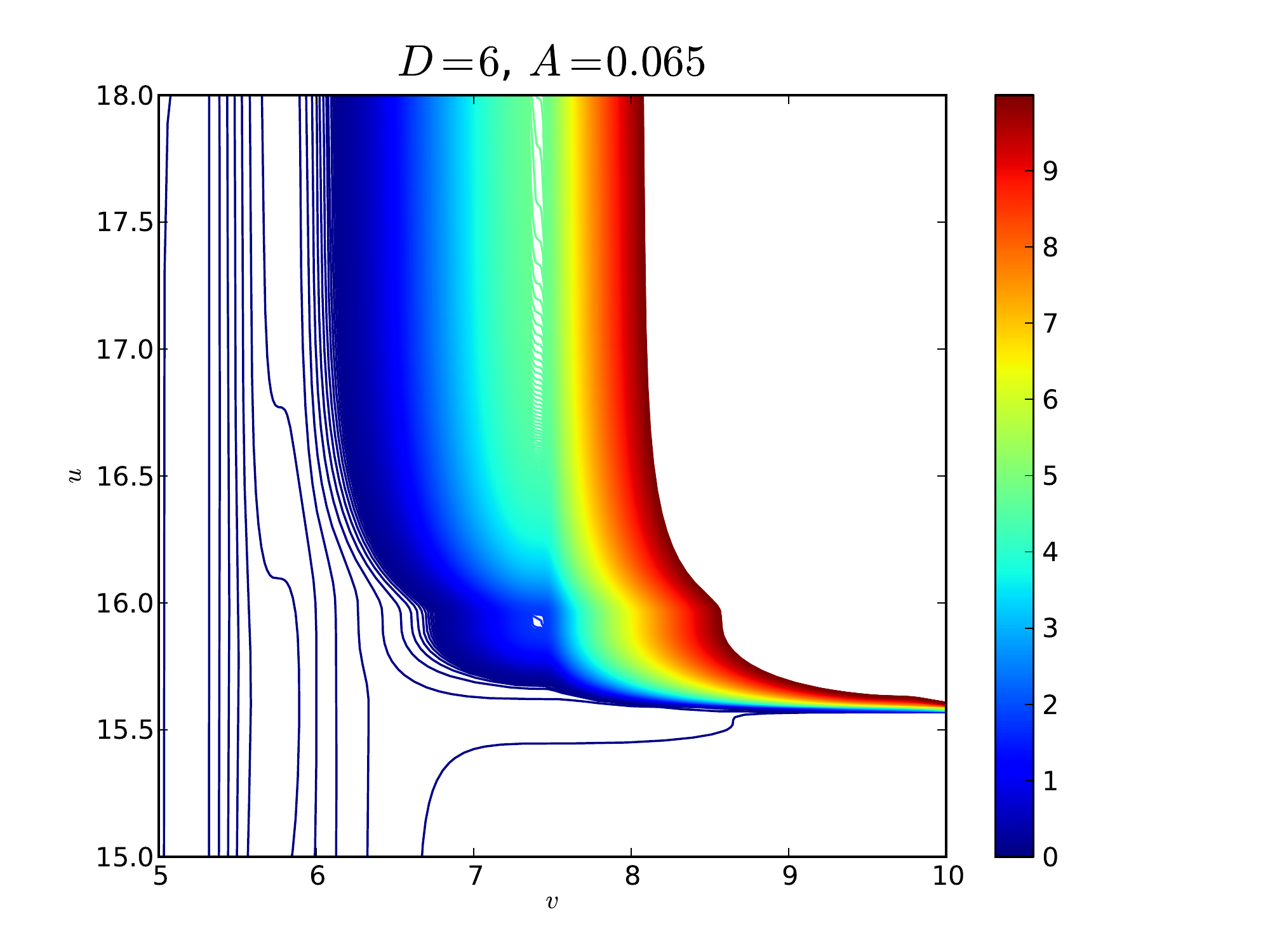}
         \includegraphics[clip=true,width=0.49\textwidth]{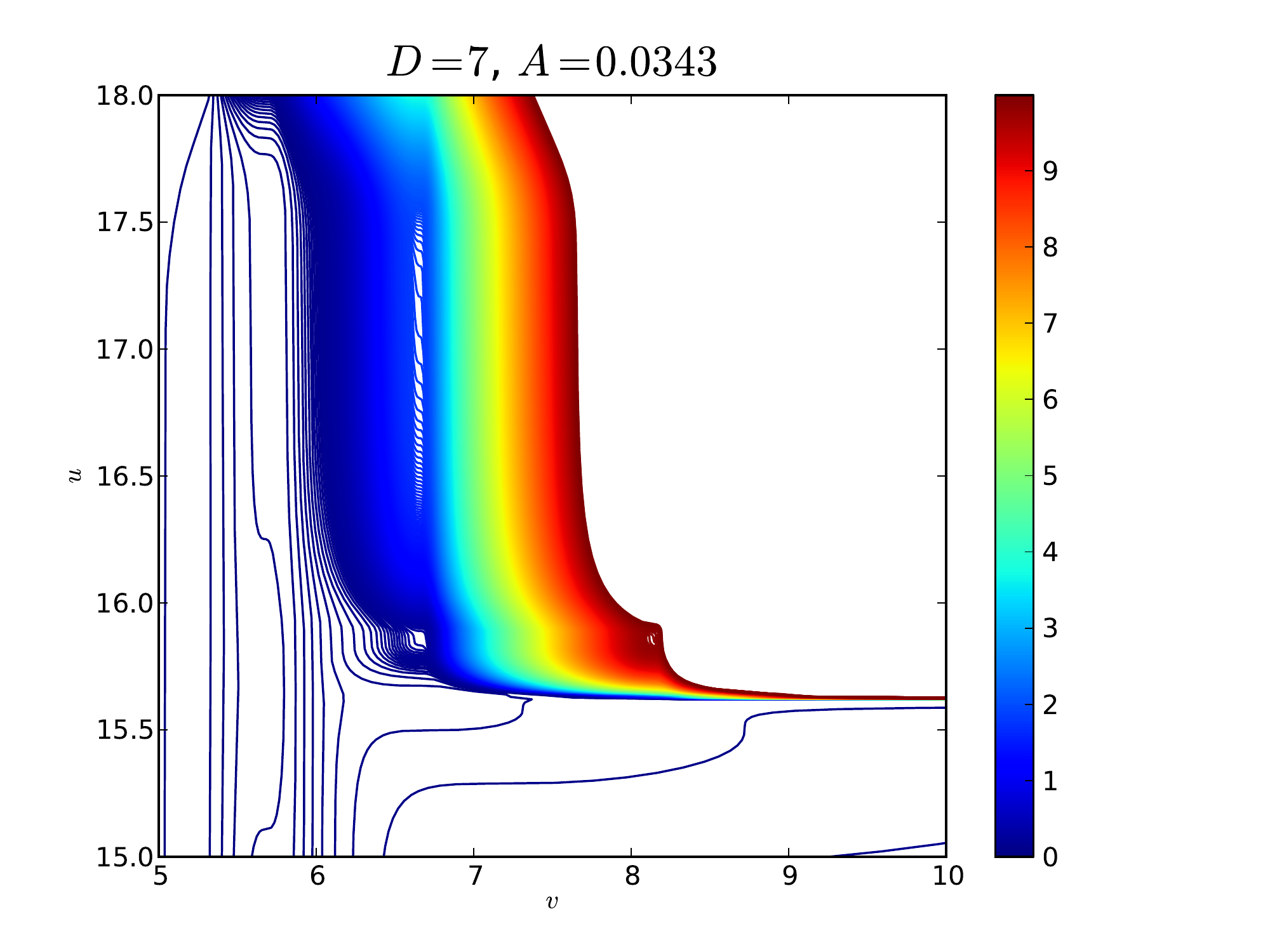}
  \caption{Misner-Sharpe mass function for a perturbed RN black hole in $D=4,5,6,7$, with the amplitude of the perturbation chosen such that the relative mass variation of the black hole is the same in all dimensions. As the dimension increases larger masses are achieved for lower $u,v$.}
  \label{mass}
\end{figure*}

In Fig. \ref{mass} we have plotted the Misner-Sharpe mass function
in the mass inflation region. The plots show that larger masses occur for lower values of advanced and retarded time, as the dimension increases. Although this is a gauge dependent statement, we emphasise that both $t$ and $r$ that define $u$ and $v$ at infinity are comparable coordinates in different dimensions (proper time of the observer at infinity and areal radius). 

\begin{figure*}[h!tb]
  \centering
    \includegraphics[clip=true,width=0.49\textwidth]{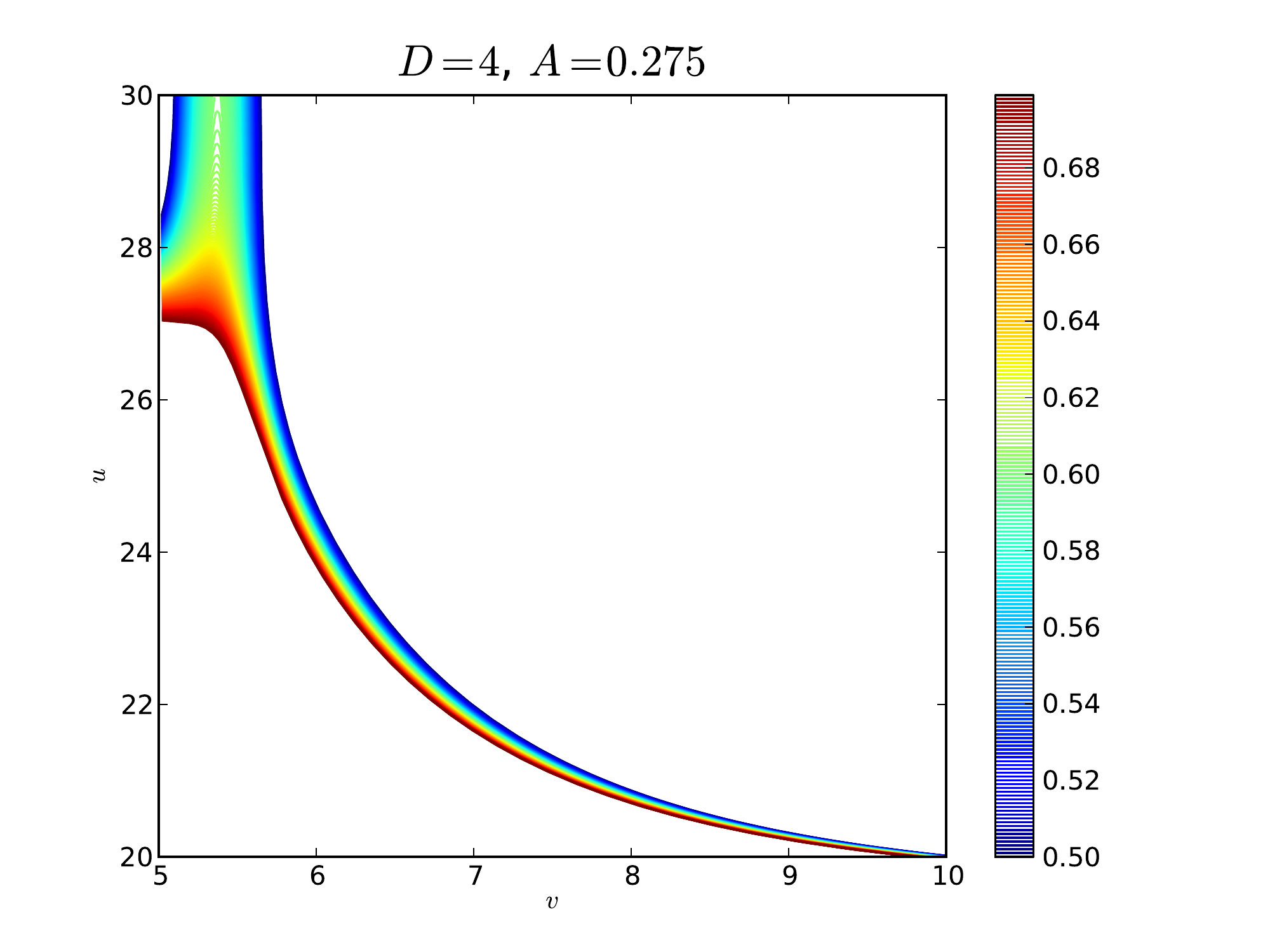}
      \includegraphics[clip=true,width=0.49\textwidth]{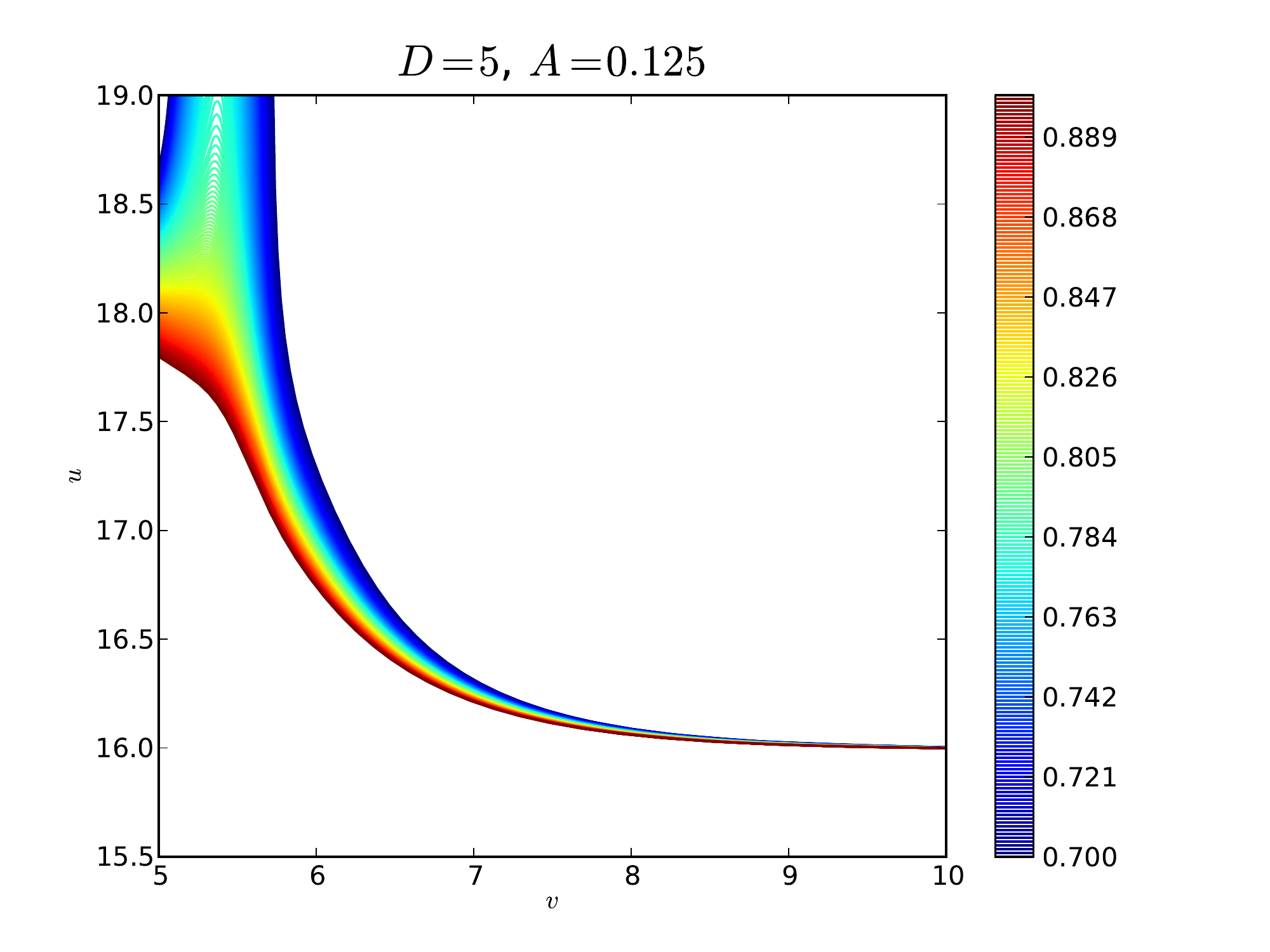}
       \includegraphics[clip=true,width=0.49\textwidth]{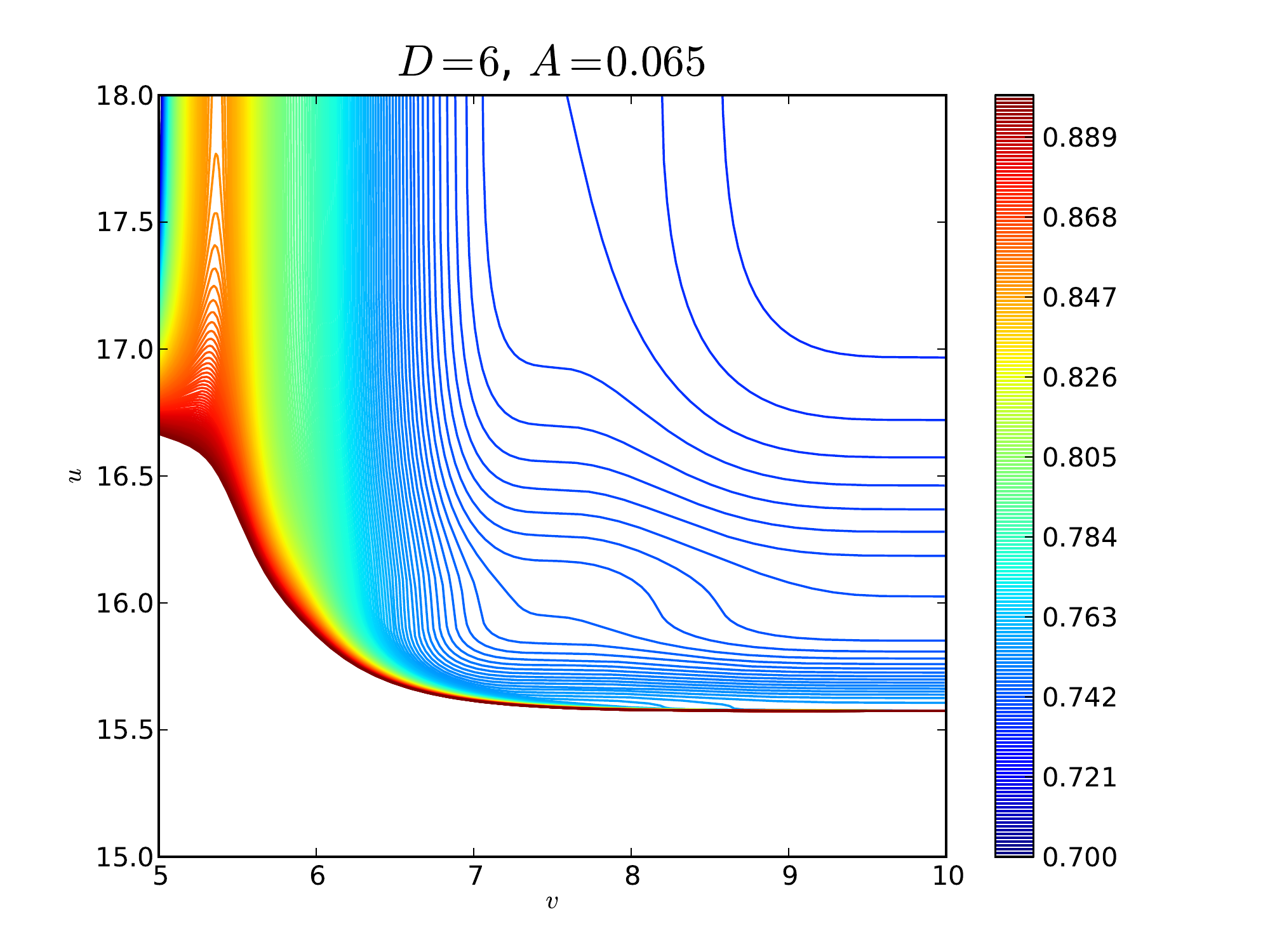}
         \includegraphics[clip=true,width=0.49\textwidth]{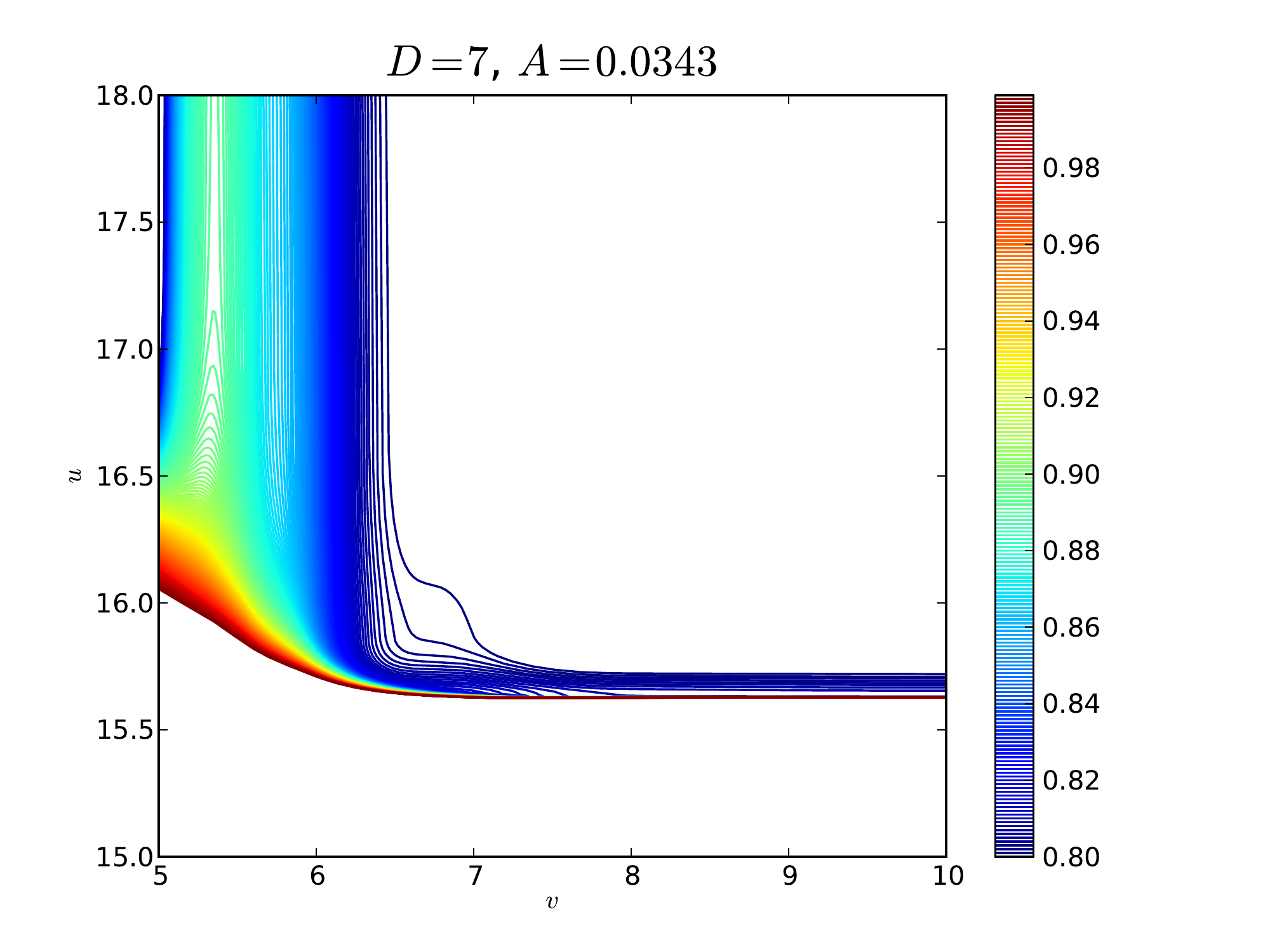}
  \caption{The radial coordinate $r$ as a function of the null coordinates $u$ and $v$ close to the inner ingoing apparent horizon for $D=4,5,6,7$ perturbed RN black hole with $M=1$ and $q=0.95$ and the amplitudes that guarantee a total mass variation of $\Delta M/M\simeq 0.25$ . Contours are plotted only for radii in the range of the color bar.}
  \label{inner}
\end{figure*}

It is tentative to conclude that the smaller time scale for the mass variations  (and also curvature as we have checked) observed in our fully non-linear numerical simulations, for increasing dimension, is consistent with the more abrupt radial variation of the metric coefficients, observed in the homogeneous approximation as the dimension increases. This is qualitatively confirmed by a simultaneous analysis of Fig. \ref{mass} and Fig. \ref{inner}, where the areal radius is plotted in the same $(u,v)$ space. Consider, for instance, a $v={\rm constant}$ line (say $v=9$), for $D=7$. From Fig. \ref{mass} it is seen that a variation of 10 orders of magnitude of the mass occurs in an interval of $\Delta u\simeq 0.1-0.2$ for $u\simeq 15.7$. Comparing with Fig. \ref{inner} we see that the variation of the areal radius in the same interval is around $\Delta r\simeq 0.1-0.2$. In this $r$ interval, therefore there is a mass variation of 10 orders of magnitude. A similar analysis, along the same $v={\rm constant}$ line in a lower $D$ shows a considerable smaller mass variation for the same $r$ interval. For instance, in $D=4$ the mass variation is of $1-2$ orders of magnitude in the same $r$ interval.

One must be careful in making this connection, however, since both $(u,v)$ and $r$ are simply coordinates and therefore gauge dependent; moreover in each case one is fixing different quantities. In particular, in the homogeneous approximation one is fixing the initial energy density at a given point, rather than the total energy in the perturbation.

\section{Conclusions}
If micro black holes form by trans-Planckian scattering collisions in the present or in a future generation of particle accelerators, they should undergo, after formation, a balding phase, in which they loose gravitational (and other interactions') multipoles and approach a stationary solution, a semi-classical phase, in which the black hole decays by Hawking radiation, with concrete and well defined signatures that could be observed in particle experiments \cite{Aad:2009wy}, and finally a mysterious (with present day physics) Planckian phase \cite{Giddings:2001bu}. 

For the scenarios proposed in  \cite{ArkaniHamed:1998rs,Antoniadis:1998ig,ArkaniHamed:1998nn,Randall:1999ee,Randall:1999vf}, these black holes are brane world black holes. But if their mass is  sufficiently above the fundamental Planck scale and their size is sufficiently smaller than the scale of the large extra dimensions, then, in the absence of charge, they should be well approximated by the classical solution for uncharged black holes in infinite $D$ dimensions. The charged case is more involved, since the electromagnetic field should propagate only on the brane (which must have a length scale smaller than the TeV) and therefore in $3+1$ rather than in $D$ dimensions. Modelling such black holes as a $D$ dimensional RN black hole is the simplest toy model, which we have explored in this paper. 

According to the result of this paper, the brane world black holes will undergo mass inflation and, potentially, form new smaller black holes inside of themselves, at least if the classical/semi-classical stage has a sufficiently long time scale. One could think this would lead to a sort of fractal, self-similar structure of black hole generations. Remember, however, that the minimum mass for the black holes is the (fundamental) Planck mass, which therefore should limit the number of generations to one.

One other possible scenario is as follows. These extremely energetic regimes are also expected to be attained in the early universe. Although the relevant physics is still not fully developed, it has been postulated that an inflationary universe might naturally arise in such extreme conditions, solving some of the problems of the standard cosmological model. Hence, the possibility that mass inflation inside black holes may also trigger the creation of new expanding universes should be considered. In \cite{Smolin:1994vb} it was assumed that quantum effects could remove the black hole singularities and be responsible for the birth of new cosmologies with slightly different values of the cosmological parameters, thus leading to a selection criteria based on the maximization of the number of black holes. The fact that Planckian and trans-Planckian regimes are attained in an extended mass inflation region much before the central singularity and the role played by the number D of spacetime dimensions on the corresponding dynamics must therefore be taken into account in such studies.

\begin{acknowledgments}
M.Z. is funded by FCT through grant SFRH/BD/43558/2008.  AJSH was supported in part by NSF award AST-0708607. This work was partially supported by FCT - Portugal through project  CERN/FP/116341/2010.
\end{acknowledgments}
\bibliography{higherD.bib}
\end{document}